%% file: heralhc.tex
\documentclass[twoside]{dis04}
\def\runauthor{J. M. Butterworth and T. Carli}
\def\shorttitle{QCD UNCERTAINTIES AT LHC AND IMPLICATIONS OF HERA}

\usepackage{rotating}
\usepackage{wrapfig}

\input jmbsym.tex

\begin{document}

\title{QCD UNCERTAINTIES AT THE LHC AND THE IMPLICATIONS OF HERA}

\author{J. M. BUTTERWORTH$^1$ and T. CARLI$^2$}

\address{
1) Department of Physics and Astronomy\\
University College London, Gower St., London, WC1E 6BT\\
E-mail: J.Butterworth@ucl.ac.uk \\
2) CERN, Experimental Physics Division, \\
CH-1211 Geneva 23, Switzerland \\
E-mail: Tancredi.Carli@cern.ch
} 

\maketitle

\abstracts{ Strong interaction physics will be ubiquitous at the Large
Hadron Collider since the colliding beams consist of confined quarks
and gluons.  Although the main purpose of the LHC is to study the
mechanism of electroweak symmetry breaking and to search for physics
beyond the Standard Model, to maximise the precision and sensitivity of
such analyses it is necessary to understand in detail various
perturbative, semi-perturbative and non-perturbative QCD effects. Many
of these effects have been extensively studied at HERA and will be
studied further at HERA II. We discuss the impact of the knowledge
thus gained on physics at the LHC.  }

\section{Introduction} 

The large hadron collider (LHC), currently under construction at CERN,
will collide protons on protons with an energy of $7$~{\rm TeV},
extending the available centre-of-mass energy ($\sqrt{s}$) by an order
of magnitude compared to existing colliders. Together with its high
collision rate, corresponding to an expected integrated luminosity of
$10-100~{\rm fb}^{-1}/{\rm year}$, these energies give access to particles with
high masses, or at high transverse momentum, which have low production
cross-sections. The LHC can search for new interactions at very short
distances and for new particles beyond the Standard Model (SM) of
particle physics.

One year of data taking at LHC energies can produce jets with
transverse energies of up to $\ETJ = 3$~TeV, probing the structure of matter
at the smallest distances ever accessed. About $20$ ($2$) $W^\pm$- ($Z^0$)-bosons will
be produced per second. These large data samples allow their
production cross-sections, and the $W^\pm$-mass, to be measured with a
precision of up to $1\%$.
Their rate provides a luminosity monitor, limited only by the
precision of the theoretical predictions.  Moreover, each second about
one top quark pair will be produced.  Precise determination of the top
quark mass and of the top decay modes will be a key challenge to the
SM.

In all these cases, a good understanding of particle production and
decay in a hadronic environment is needed.
The large phase space and the large cross-section of strongly
interacting particles makes it necessary to test and extend our
understanding of strong interactions in the early phase of LHC data
taking.  This requires not only an experimental program to measure
basic SM processes over a wide range, but also the development of
``tuneable'' models implementing the correct underlying physics
processes, and of tools to calculate higher order cross-sections.
Examples are models for soft hadron-hadron collisions (minimum bias
and underlying events), simulation of events with many particles and
many jets and their correlations, and the production mechanism for weakly
interacting particles and heavy quarks.  It will also be necessary
to validate the QCD input parameters, the strong coupling constant and
the parton density functions, extrapolated to high momentum transfers,
and to constrain their uncertainties with LHC data.

Thus, studying QCD in high energy collisions is interesting now, and
also vital for the future of high energy physics. The ability of the
Tevatron and earlier hadron-hadron machines to have an impact in this
area is clear, and is the subject of ongoing study~\cite{tev4lhc}. The
focus of this contribution, however, is the impact of data from the
HERA lepton-proton collider at DESY. HERA is a precision QCD machine,
as well as a QCD ``discovery" machine. We argue that data from HERA
are needed to fully exploit the LHC. The areas we discuss can be split
into three general categories: the precision measurement of QCD input
parameters; the testing of calculational techniques; and the testing
of non- or semi-perturbative models.  In all these areas, HERA data
can help us gain a quantitative understanding of hadronic production
mechanisms at high energies.

\section{Precision measurement of QCD inputs}

\subsection{The Strong Coupling Constant}

\begin{wrapfigure}[19]{l}{7.5cm}
\vspace*{6.5cm}
\begin{center}
\includegraphics{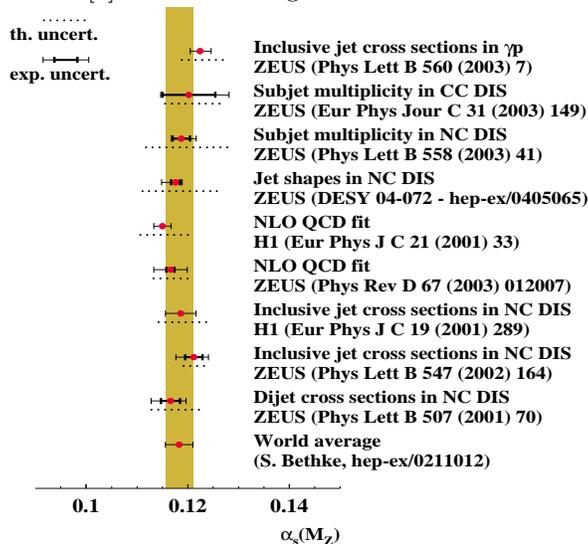}
\vspace*{-0.3cm}
\caption[*]{ 
$\alpha_s$ measurements from HERA I~\cite{alphas}.
\label{fig:as}}
\end{center}
\end{wrapfigure}

One yardstick by which the present understanding of the strong
interaction can be judged is the precision of measurements of the
strong coupling constant $\alpha_s$, the fundamental parameter of
QCD. A summary of such measurements made at HERA~\cite{alphas} is
shown in Fig.~\ref{fig:as}.

The majority of the measurements shown are made using jet cross-sections 
and event shape properties in the final state.  To make them,
many technical advances in QCD calculations have been exploited, and a
calibration of the calorimeter energy response to around the 1\% level
was required. 
The parameter $\alpha_s$ itself is perhaps not such a
critical number for the LHC. 
However, measuring $\alpha_s$ at the high scales
uniquely accessible at the LHC is an essential measurement which
probes QCD at very small distance scales and is hence sensitive to
physics beyond the SM (see also Section \ref{sec:pdf}).

The HERA measurements demonstrate the
ability to do precise QCD physics in the final
state of hadronic collisions. This ability has a significant
impact in several areas of more direct relevance to the LHC.

\subsection{Parton Distributions}
\label{sec:pdf}

The most obvious area where HERA data have an impact at the LHC is the
parton distribution functions (PDFs) in the proton. These
distributions parameterise the probability of resolving, at a given
energy scale $\q2$, a quark or gluon in the proton carrying a fraction
x of the proton's momentum. They are thus crucial inputs to all hard
cross-section predictions at the LHC. They are determined using fits
to deep inelastic scattering data and other processes; the HERA
data~\cite{heraf2} are a dominant input to these fits, particularly in
driving the rise of the gluon density as x decreases.

\begin{figure}[!thb]
\vspace{11.0cm}
\begin{center}
\includegraphics{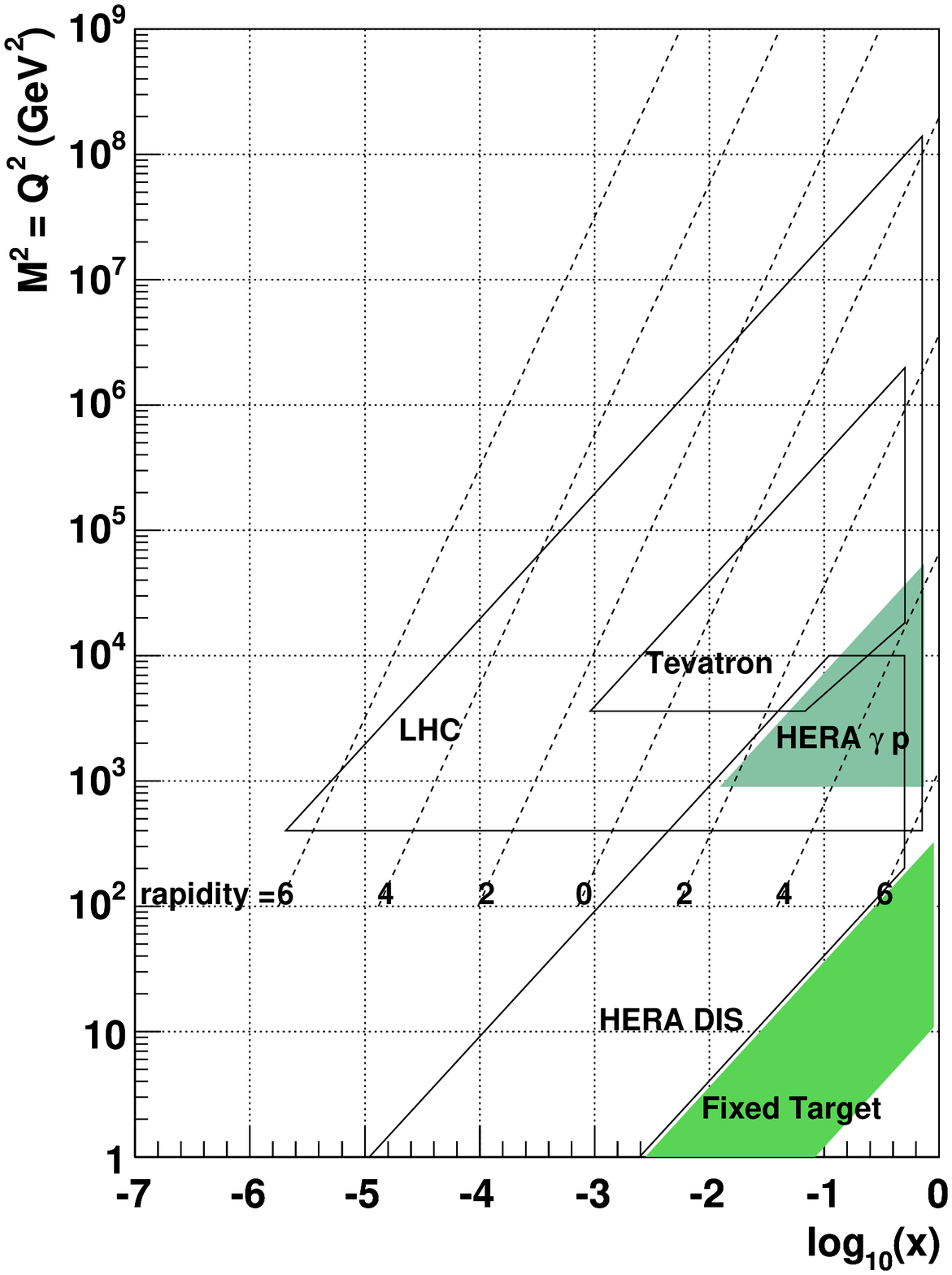} 
\includegraphics{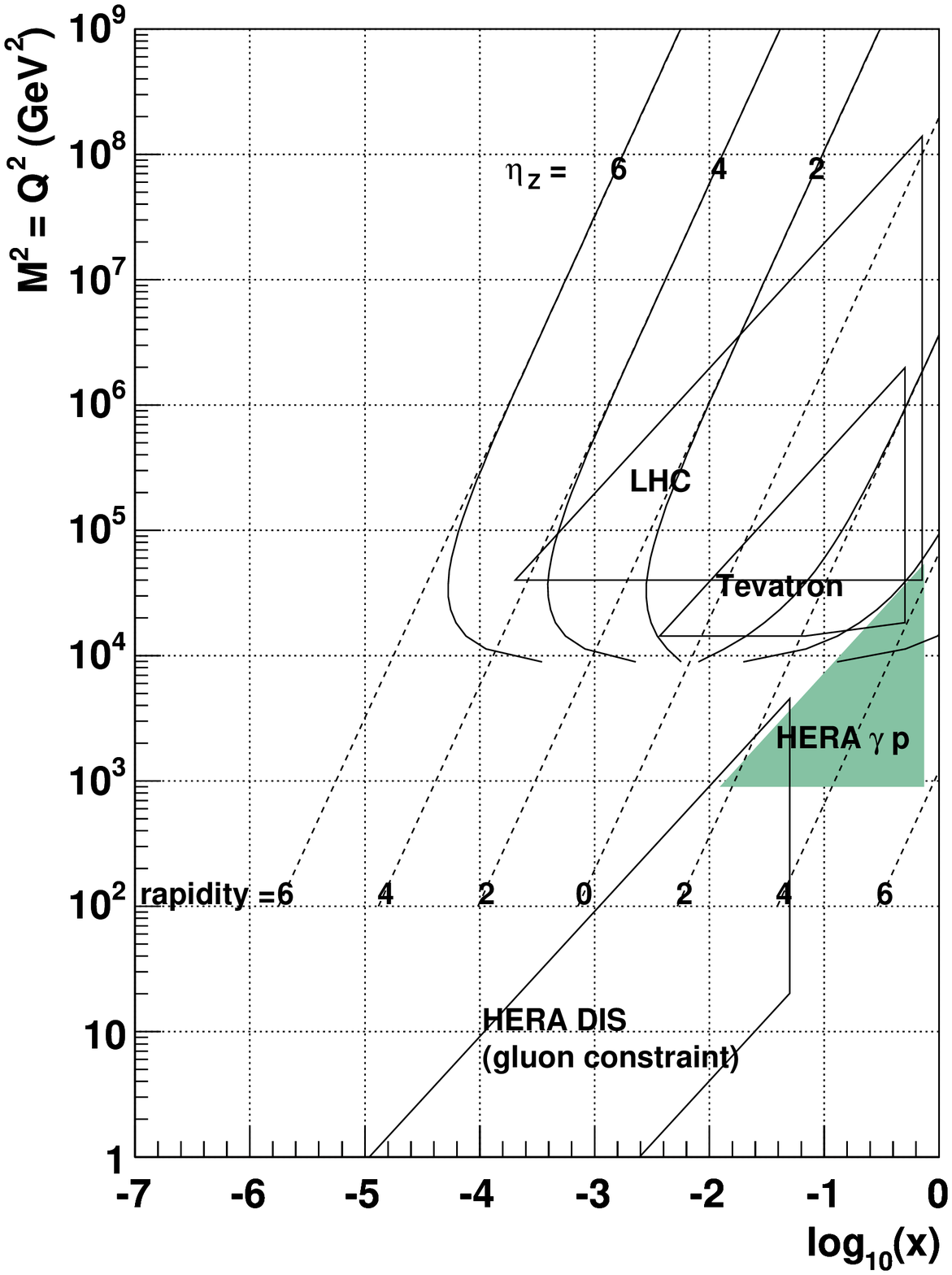} 
\begin{picture}(0,0) 
\put(-150,15){a)} 
\put(20,15){b)}
\end{picture}
\vspace{-0.5cm}
\caption[*]{ The parton kinematic plane. In (a), the approximate
region of LHC sensitivity to the PDFs is shown, along with the regions
where pre-LHC measurements either constrain or are expected to
constrain them. In (b), the regions in which LHC, HERA and Tevatron
constrain (or can constrain) the gluon PDF are shown. The assumption
is that jets below $100$~{\rm GeV} ($60$~{\rm GeV}) at LHC (Tevatron) cannot be
used to get precise constraints. The diagonal lines show the rapidity
of the system with mass $M$ at LHC, and the curves show the
approximate corresponding pseudorapidity for a $Z$ boson produced at
such a rapidity (see text). The HERA photoproduction region is taken
from~\cite{zeusgp}.
\label{fig:kp}}
\end{center}
\end{figure}

However, when one translates the parton kinematics to the x, $\q2$
plane (see the well known illustration in~\cite{Martin:1999ww}, and
Fig.~\ref{fig:kp}a) it is clear that there is in fact only a small
overlap between the HERA and LHC regions. The DGLAP~\cite{DGLAP}
equations are therefore used to evolve the parton densities up in
$\q2$ and obtain predictions for the LHC.  In addition the LHC, by profiting
from the techniques developed for the hadronic final state analyses at HERA,
will itself be able to constrain PDFs using a variety of SM processes.

The limitations due to the present PDF uncertainties can be
illustrated with the example of inclusive jet production at
LHC. The measurement of the single inclusive jet cross-section as a
function of $\ETJ$ is a clean way to demonstrate our understanding of
$\alpha_s$ and of the PDFs at very short distances, and in addition a
promising way to search for new physics.
At central rapidity $0< |y| < 1$ an uncertainty of $100 \%$ is found
at $\ETJ \approx 5$~TeV. For forward jets $2< |y| < 3$ this large
uncertainty, which is mainly due to the limited knowledge on the gluon
distribution at large x, is already present at $\ETJ \approx
2$~TeV~\cite{cteqjets}.  By measuring the inclusive jet
cross-section in different rapidity bins, it should be possible to
disentangle possible new physics from other, more mundane,
effects. However, the PDF uncertainties do reduce the ability of the
LHC experiments to discover new physics in what is a relatively simple
channel, where due to the inclusive nature of the measurement the
background calculations should be in principle very reliable.

For instance, the dijet cross-section are sensitive to possible
effects due to extra space-time dimensions. In contrast to the SM,
where the electroweak symmetry breaking scale, the GUT scale and the
Planck scale are very different, models with extra space-time
dimensions, compactified at a scale $M_c$, need only one fundamental
scale, which may be of the order of a few {\rm TeV}. If bosons can
propagate in these extra-dimensions, a modification of the energy
dependence of the strong coupling is expected \cite{EXTRADIM}.  In
principle the dijet cross-sections at LHC give a sensitivity to
compactification scales up to $5 - 10$ {\rm TeV} \cite{Balazs:2001eu}.
However, PDF uncertainties reduce this sensitivity to $2 - 3$ {\rm
TeV} \cite{Ferrag:2004ca}.

There remain several issues with the above: 
\begin{enumerate}

\vspace{-0.1cm}
\item
The HERA and fixed-target data are primarily sensitive to the quark
distributions; the gluon is at least equally important at the LHC. The
gluon is constrained in the global fits mainly via scaling violations.
However, this means that if the quark is measured in some x range,
this will determine the gluon distribution in a region of lower
x. Deep inelastic scattering data provide little effective constraint
on the gluon for x $\ge 0.1$ (see Fig.~\ref{fig:kp}b), and there are
significant uncertainties even below this. This situation will not be
improved by any measurement of $F_L$ at HERA II~\cite{max}. Some
information can be obtained from jet data at the Tevatron, but the
precision is not competitive with structure function
measurements. Furthermore, some of this data represents the highest
energy parton-parton collisions ever measured, and new physics
possibilities cannot be excluded. The net result is - the knowledge of
the gluon density at high and intermediate x is poor.

\vspace{-0.1cm}
\item
The LHC will have difficulty constraining the parton distributions at
high x. At the highest $\q2$, the problem is the same as that
currently faced at the Tevatron - these will be the highest energy
parton-parton collisions ever, where new physics may arise (see
above). What is required is a constraint at intermediate $\q2$ and
high x. To achieve this, it is necessary to use boosted events, where
one x value is high and the other low. If one considers $Z$+jet
production in the region of interest, say at rapidity of 2 and a
$Z$+jet mass of $600$ {\rm GeV}, the x values probed are $3.2 \times 10^{-3}$
and 0.58.  The rapidity and pseudorapidity lines for $Z$ bosons are
shown on Fig.~\ref{fig:kp}b. The detector acceptance will drop between
pseudorapidities of 3 and 4. Detailed MC studies are required to see
where the LHC will really provide good information.

\vspace{-0.1cm}
\item
It is far from certain that DGLAP evolution is valid over the required
x range. The current fits show good agreement down to x $\approx
10^{-4}$, but the amount of data at low x for $\q2 \gsim 4~\gev2$,
where fits can be made, is small. Hence the level of confirmation of
the validity of the evolution in this region is weak. The low x parton
in the example above (x $= 3.2 \times 10^{-3}$) is already in a region
where a more stringent validation of the applicability DGLAP evolution
is desirable before predictions are made with confidence.  Here, more
measurements at HERA, including $F_L$, 
together with charm and beauty quark cross-section measurements in DIS and
photoproduction, as well as measurements in the early days of LHC, will
be very important.
\end{enumerate}

\begin{figure}[!thb]
\vspace{11.0cm}
\begin{center}
\includegraphics{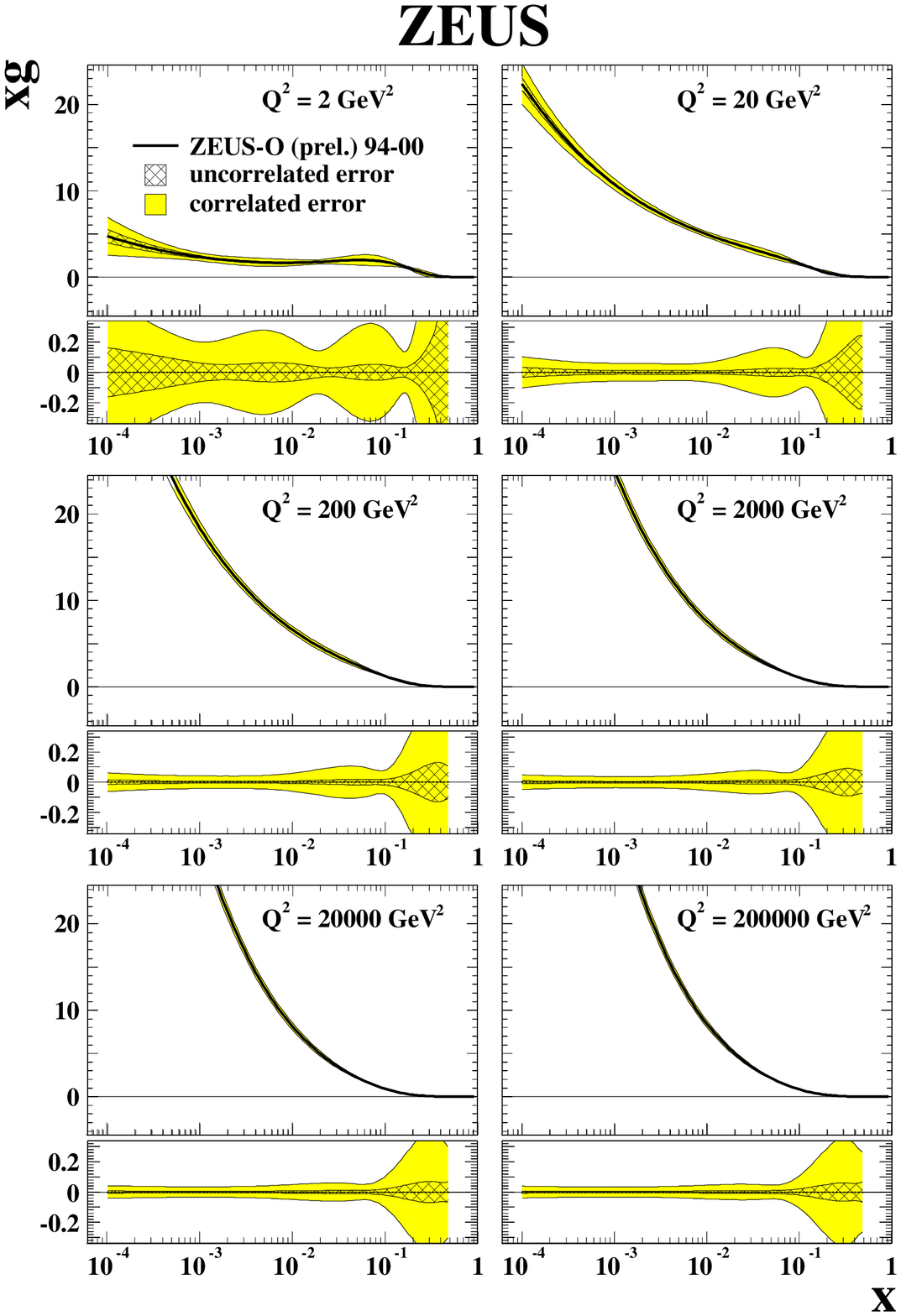} 
\includegraphics{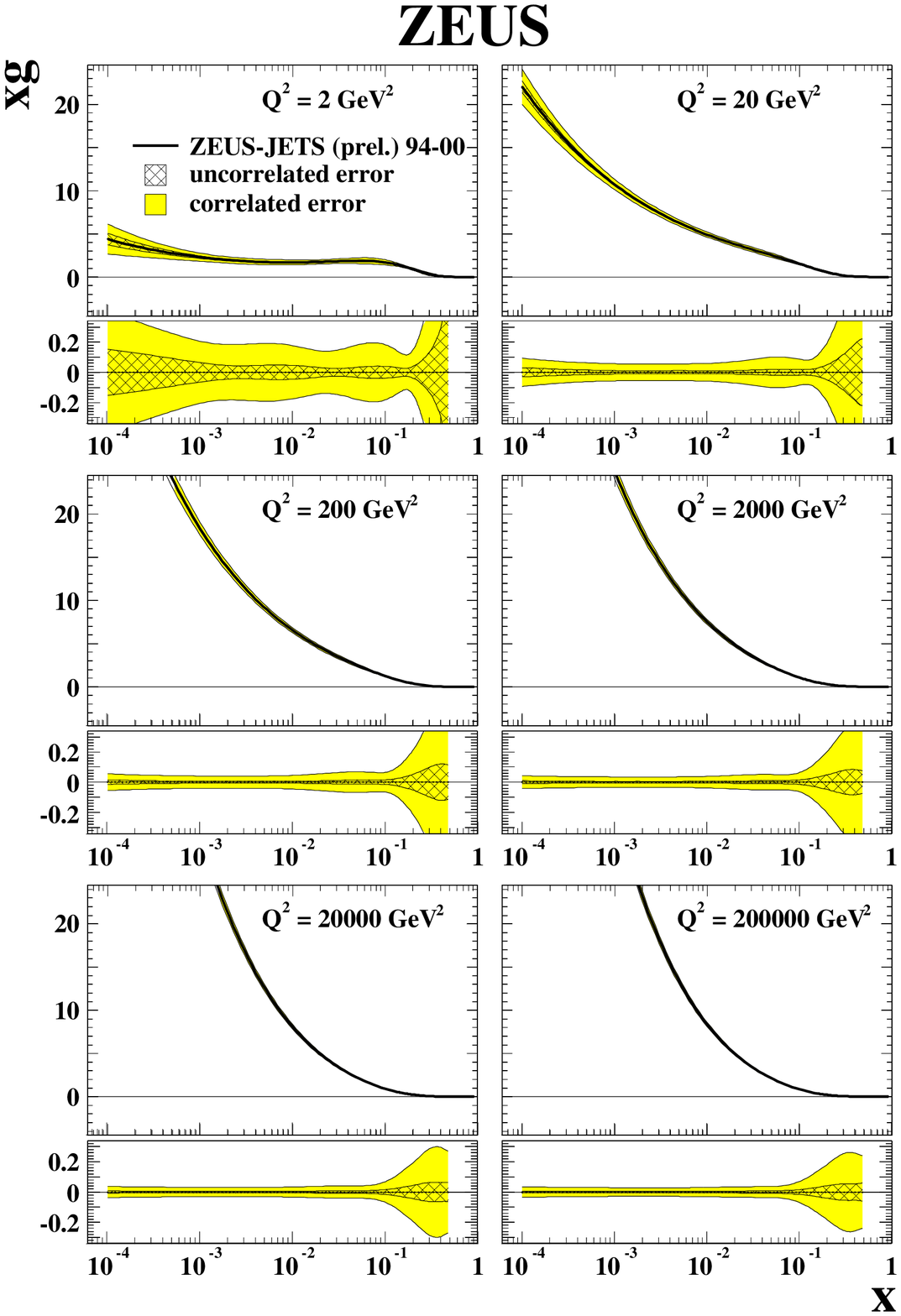} 
\begin{picture}(0,0) 
\put(-160,15){a)} 
\put(20,15){b)}
\end{picture}
\vspace{-0.5cm}
\caption[*]{ ZEUS fits~\cite{mandy}. In (a), only ZEUS DIS jet data
 are used.  In (b), ZEUS jet data~\cite{zeusgp,zeusdisjets} are added
 to the fit. The yellow band shows the total uncertainty, which at
 high x and high $\q2$ is reduced from around 40\% to around 25\%.
\label{fig:zeusfit}}
\end{center}
\end{figure}
\begin{figure}[!thb]
\begin{center}
\includegraphics{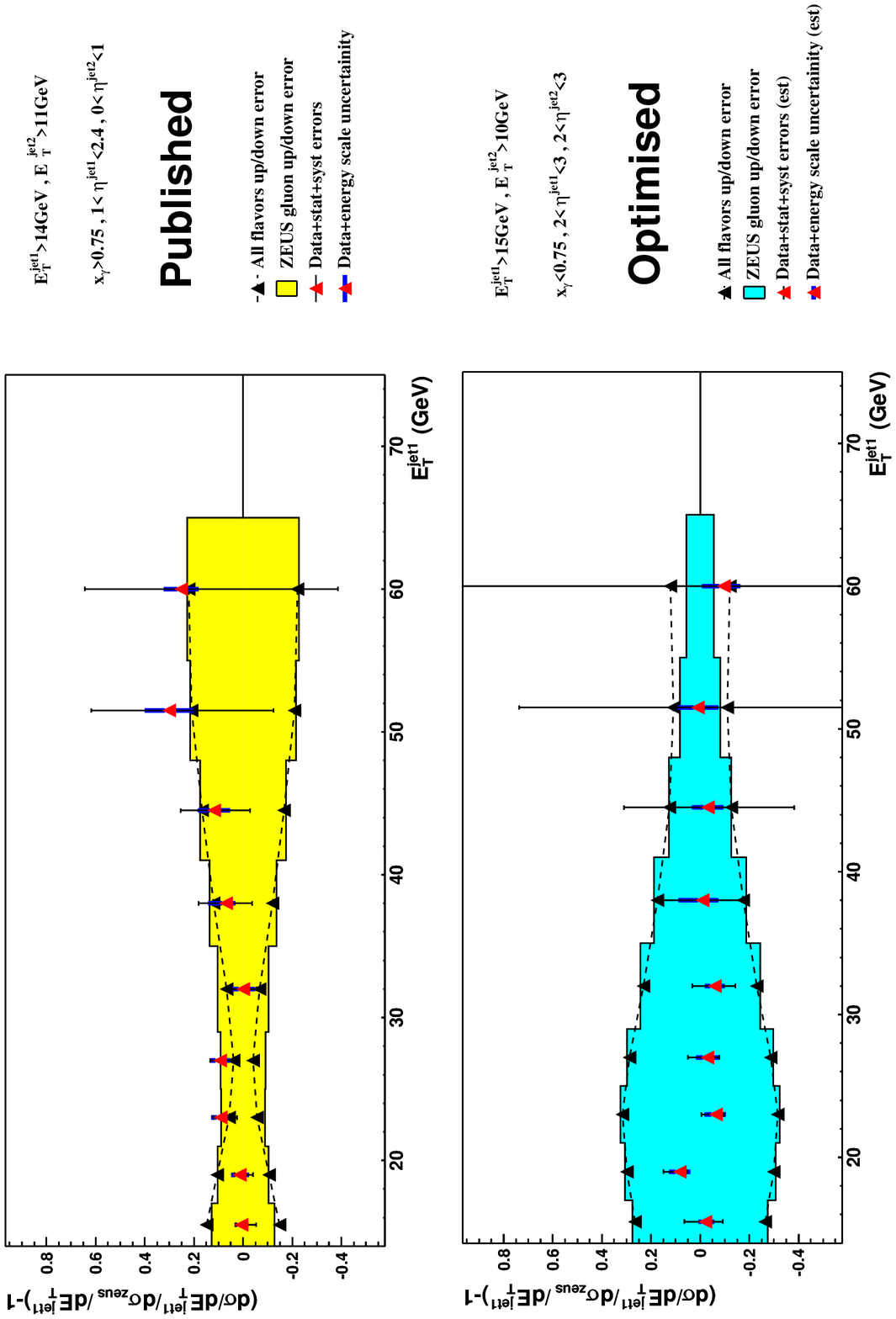} 
\vspace{10.0cm}
\caption[*]{ The potential of HERA II jet photoproduction
data~\cite{cta}. The upper plot compares the uncertainties on the
ZEUS-O~\cite{mandy} fit to the relevant published jet photoproduction
data~\cite{zeusgp}. The lower plot compares the uncertainties from the
same fits, but now for a cross-section optimised to constrain the
gluon. The data in the lower plot are faked, but give an indication of
the experimental uncertainties expected for $500~pb^{-1}$ of HERA II
data.
\label{fig:zeusjets}}
\end{center}
\end{figure}

It should be obvious that anything more HERA can say about the high x
region will be very valuable. One under-exploited process with
potential to help here is dijet photoproduction. The kinematic reach
of this process, which is directly sensitive to the gluon density, is
shown in Fig.~\ref{fig:kp}. The ZEUS collaboration has already
included jet data (from DIS and
photoproduction~\cite{zeusgp,zeusdisjets}) in their new
fits~\cite{mandy}. The impact is considerable, particularly in the
region x$ > 10^{-2}$ (Fig~\ref{fig:zeusfit}). The data set used is
ZEUS 1996-1997 data, which is statistically limited at high $\ETJ$,
corresponding to high x. Several possibilities for improvement include
optimising the cross-section for sensitivity to the high x gluon by
(for example) extending to the forward region, including the rest of
ZEUS and H1 HERA I data, and eventually HERA II data. An illustration
of the potential of HERA II data in this area is shown
Fig.~\ref{fig:zeusjets}. It is important that these possibilities are
vigorously pursued, not just by the collaborations, but by the other
global fitting groups.

Another area where there has been impressive progress recently is in
the full calculation of the DGLAP splitting functions to
next-to-NLO~\cite{nnlo}. These calculations show that the perturbative
series is stable over a wide kinematic range (for $x > 10^{-3}$ the
NNLO corrections are around eight times smaller than the NLO
terms). This gives more confidence and accuracy in the use of the
evolution from the HERA regime, and (coupled with NNLO matrix elements
for key process) will allow the LHC to more accurately constrain the PDFs.

\subsection{Fragmentation Parameters}

Another set of parameters which may be measured at HERA and used at
LHC are those governing fragmentation, in particular the one
of charm and beauty quarks.

For example, HERA can measure the charm fragmentation function
directly in hadronic events~\cite{sanjay}. The same should also be
possible for $b$-quarks with HERA II data, and precise measurements of
such parameters would be an important input to predictions for $b$-jet
rates, and important process at LHC.  The fragmentation fractions for
charmed particles measured at HERA also measured with precision
comparable to the combined LEP results, and give confidence in the
portability of such numbers between different processes~\cite{leonid}.

\section{Testing ground for calculational techniques} 

The LHC will operate at energies where processes with multiple hard
scales ($M_W, \Mt, \ETJ$) are commonplace. Such an
environment is a new challenge for the established techniques of
perturbative calculation. At HERA, the scales involved are lower
(e.g. $Q^2, \ETJ, M_c, M_b$) but the experiments have unique control
over them and the data provide an equally challenging arena for QCD
predictions.

\subsection{Very forward jets}

At the LHC, forward jets ($2 < \ETAJ < 4$) are of interest for several
reasons. In general, extending the acceptance as far forward as
possible will enhance search channels, and is also important for
determining the PDFs (see Section~\ref{sec:pdf}). They may also be
useful for triggering purposes at LHCb~\cite{lhcb}. But perhaps the
most important application is their role in tagging vector-boson
fusion events~\cite{tagjets}, a key SM process and Higgs
search channel.

Forward jets are also sensitive to low-x
physics~\cite{mn}. Although the jet itself is at high-x, the QCD
evolution between two high-$\ETJ$ jets at widely differing rapidities
is sensitive to non-DGLAP evolution~\cite{bfkl}. A similar kinematic
configuration is achieved in forward jet production in DIS at
HERA~\cite{herafj}. There has been much study of these processes at
HERA~\cite{fjmeas}, and a general feature is that the agreement with
fixed order QCD calculations degrades, and the theoretical
uncertainties in such calculations increase, as the jet moves into
the forward region. This may be a sign that low-x effects are becoming
important. It certainly means that predictions of rates at LHC need
careful study. Fixed-order calculations of the signal~\cite{zep2} show
that the uncertainties are around the 5\% level. However, the
uncertainties from QCD in the backgrounds to these processes may be
much higher and will be hard to evaluate. They may be a dominant
effect in the measurement of the $WW$-fusion cross-section. Studying
such effects at HERA is therefore important for improving the
phenomenology in this area.

\subsection{Multijets}
In many searches for physics beyond the SM, multi-jet events are an
important background, {\it e.g.} for SUSY searches. The correct
modeling of the correlation between the jets is therefore crucial to
maintain the optimal sensitivity of the LHC experiments. Another
example where the multi-particle final state has to be correctly
modeled is the production of top quark pairs at LHC. One of the most
promising channels for the accurate determination of $\Mt$ has
four jets in the final state: $p p \to t_1 t_2 \to W^\pm W^\pm b_1 b_2
\to b_1 \nu_l l b_2 q q$.  Using the PYTHIA and HERWIG simulation
programs, the modeling of the higher order final-state parton
radiation is expected to give the largest systematic
error~\cite{topmass}.  It limits the error on $\Mt$ to about $1$ {\rm
GeV}.  In view of the potential to discover new physics through a
mismatch in the correlation of the $W^\pm$, the top quark and the
Higgs mass, efforts to improve our understanding and ability to
simulate higher order QCD radiation have recently intensified.

These efforts mainly focus on simulation of the final state in $pp$
collisions via Monte Carlo (MC) simulations and/or automatic resummed
calculations~\cite{caesar}. One of the most important developments is
the consistent implementation of higher order parton radiation in full
MC event generators.  Here, two different possibilities have been
explored: the matching of NLO matrix elements (ME) to parton showers
(PS) in programs like MC@NLO \cite{MCNLO,frix} and the matching of
$n$-parton tree level ME with PS.

In the MC@NLO approach the analytic form of the real parton emissions
and the virtual corrections is subtracted from the NLO ME. Then the
singularity connected to soft and collinear parton emissions is
handled by the PS algorithm and absorbed in the Sudakov form
factor. The NLO corrections are used to describe the $n$-body
kinematics. Since negative and positive event weights are bounded, an
unweighting procedure can be performed.
This procedure has been successfully implemented for many processes,
including heavy quark, Higgs, Drell-Yan and $W^\pm$ and $Z^0$
production. It can in principle be extended to any process known to
NLO accuracy. The inclusion of NLO corrections guarantees that the
total cross-section generated by the MC has the correct normalisation,
and its renormalisation- and factorisation-scale dependencies are reduced.

This procedure leads to a correct description of
the one extra parton from the Born LO processes.  If fails, however,
for events with high jet multiplicities, since hard radiation at large
angles is suppressed by the angular ordering in the PS.  To solve this
problem the $n$-parton tree-level ME can be merged with the PS.
Double counting can be avoided with the CKKW prescription \cite{CKKW}.
This approach is being implemented in the next generation of MC event
generators \cite{MCNEXT}. Presently, ME up to $2 \to 8$ parton processes
can be treated.

\begin{wrapfigure}[32]{l}{5.5cm}
\vspace*{10.0cm}
\begin{center}
\includegraphics{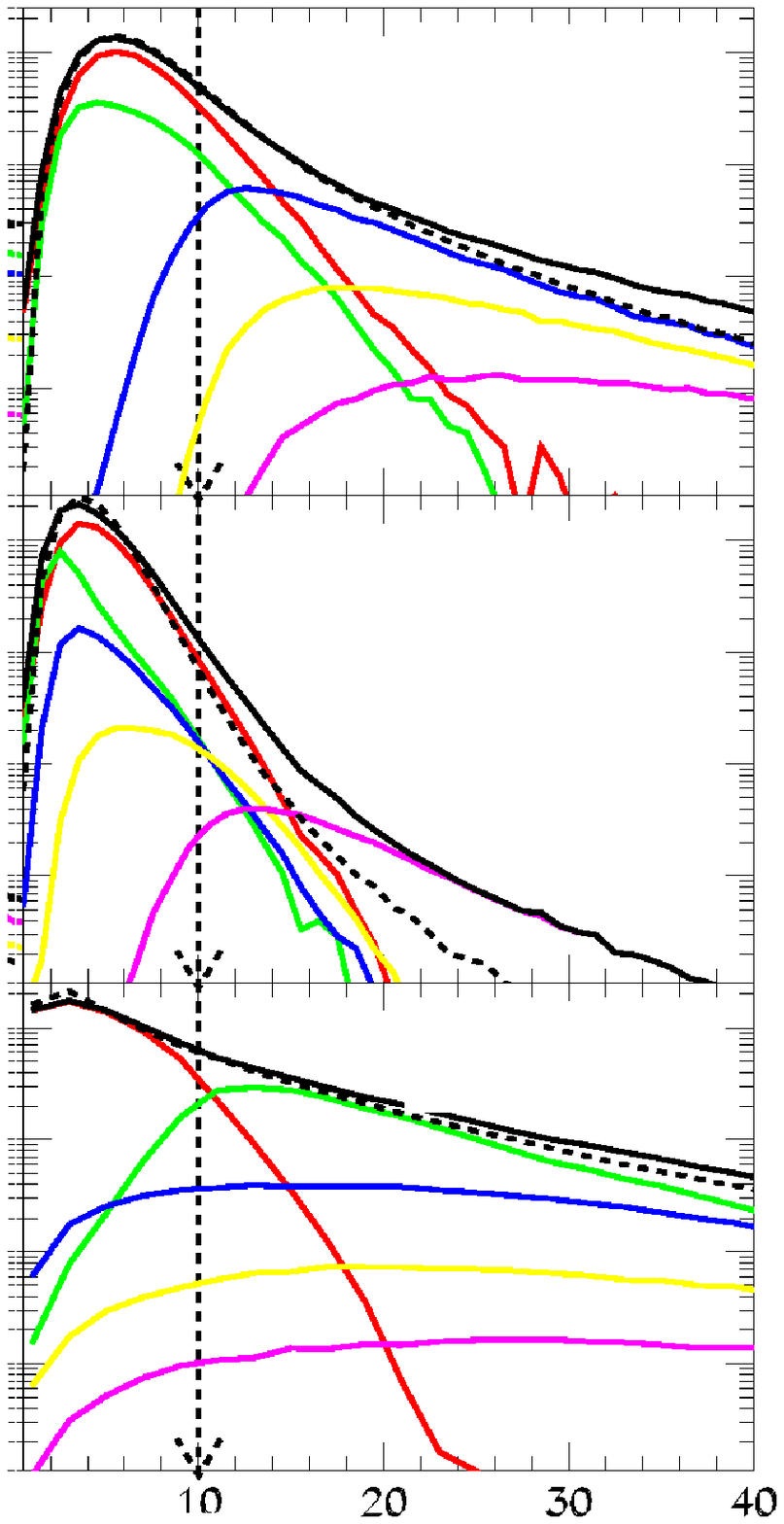}
\begin{picture}(0,0) 
\begin{rotate}{90} 
\put(150,80){$d\sigma/dK_T$ [pb/GeV]} 
\end{rotate}
\put(0,10){$K_T$ Cluster [GeV]} 
\put(30,280){$K_{T,2-jet}$} 
\put(30,190){$K_{T,4-jet}$} 
\put(30,105){$K_{T,W^\pm}$} 
\put(50,215){a)} 
\put(50,130){b)} 
\put(50,35){c)} 
\end{picture}
\vspace*{-0.5cm}
\caption[*]{\label{fig:multijets}
Differential jet cross-section for $W^\pm$ and $n-jets$ events
as a function of the transverse energy of the second (a) and forth (b) jet
and of the $W^\pm$ (c). The result of a matching
procedure of the $n$-parton tree level MEs and PS
is shown as solid line. 
The dotted line shows the HERWIG default. From \cite{JHEP0405}.
}
\end{center}
\end{wrapfigure}

As an example \cite{JHEP0405}, we show the matching of PS to the
$n$-parton tree level ME for events with $W^\pm$-bosons and $n$-jets
in $pp$-collisions at $\sqrt{s}=1.96$ {\rm TeV}.  The differential jet
cross-section with respect to 
the $K_T$-jet cluster algorithm in hadron-hadron collisions
\cite{NPB406} is shown in Fig.~\ref{fig:multijets}. The matching has
been performed at $K_T = 10$ {\rm GeV} (vertical dotted line).  The
quantity $K_T$, closely related to the transverse momentum, is shown
for the second (a) and the fourth (b) highest $K_T$ jet and the
$W^\pm$-boson (c) for the matching of the $n$-parton tree level ME
with the PS\footnote{ Also shown as coloured solid lines are the
individual $n$-parton ME components: $2$ partons as red, $3$ as green,
$4$ as blue, $5$ as yellow, $6$ as magenta solid line.} and for the
default HERWIG ME matching procedure \cite{herwigmatching} (dashed
line).  For low $K_T$, the new matching procedure and the HERWIG
default agree.  Towards high $K_T$ the new matching procedure leads to
a significantly harder distribution. This is quite dramatic for the
forth jet where the $6$-parton ME gives the main
contribution. Obviously the PS as implemented in HERWIG is not a good
approximation in this region.  Since at LHC there is a much larger
phase space for parton radiation, the correct simulation of
multi-parton final states will be even more important.

Since the above procedure is based on a LO calculation, the absolute
normalisation is arbitrary and has to be determined from the
data. However, the shape of the distributions and the correlation
between the final state particles should be well described.  For
multi-particle final states these calculations should also be better
than NLO calculations of lower multiplicity diagrams.

The validation of these ideas using Tevatron data is presently
on-going.  For HERA, these new ideas have not yet been made
available. However, the clean event topology, the controllable
kinematics and the large phase space for hadron production in the
low-x regime mean that HERA data could make decisive contributions in
this field. For instance, it was the first time at HERA that
NLO corrections for $2 \to 3$ parton processes have been calculated
for collisions with hadrons in the initial state \cite{Nagyep}.
After the successful experimental test \cite{h13jet} of the 
$3$-jet cross-sections $e p \to j j j$ in DIS it has also
been made available for proton proton collisions $p p \to j j j$ \cite{Nagypp}.
A number of other processes~\cite{3nlo} have been made available since then:
$p p \to V + j j $, 
$p p \to \gamma \gamma j$, 
$p p \to H j j$. 
These calculations represent the current frontier of our capabilities.

Furthermore, comparisons of fixed-order ME programs and
leading-logarithmic PS simulations to current HERA jet data
already show interesting features. For example, the dijet cross
section for hadronic photon events as a function of the leading jet
transverse energy is in excellent agreement with NLO QCD
calculations~\cite{zeusgp,h1gp}. There is also good agreement with
HERWIG, if a normalisation factor of 1.6 is applied. However, these
dijet cross-sections are defined in terms of highest $\ETJ$ jet and
the rapidities of the two highest $\ETJ$ jet.  If the $\ETJ$ of the
second jet is varied, this is formally a sub-leading effect. However,
there is a significant dependence of the cross-section on this
variable, and the shape of the dependence is well modeled by HERWIG,
but not by fixed order NLO~\cite{zeusgp}. Furthermore, three-jet 
cross-sections for $M_{\rm 3 jet}>50$~GeV~\cite{3j}, which are sensitive to colour
coherence in initial and final state radiation, are described by
HERWIG in both shape and normalisation if the same factor of 1.6,
determined from the high $\ETJ$ dijet data, is
applied~\cite{jetweb}. Such studies indicate the successes and
limitations of the technology and should be extended to test the newer
techniques described above.

\subsection{NNLO calculations}
NLO QCD calculations are able to describe $2 \to 2$ and $2 \to 3$ processes
in most phase space regions in both $pp$ and $ep$ collisions.
However, the predictions are in many cases still insufficient for precision
analyses. Typical residual scale dependences on the renormalisation and
factorisation scales are of the order of $10-20 \%$. 
This uncertainty on the absolute cross-section normalisation
presently limits, e.g., the precision extraction of $\alpha_s$ at hadron colliders.

At LHC, about $10^{5}$ $W^\pm$-boson and about   $10^{4}$ $Z^0$-bosons with transverse
momenta bigger than $400$ {\rm GeV} will be produced for an integrated luminosity
of $30~{\rm fb}^{-1}$. This huge data sample can be used for detector calibration
and to monitor our understanding of $pp$ collisions.
One of the obvious applications is the quasi-online measurement of the
LHC luminosity needed for the determination of couplings within the SM (e.g. $\alpha_s$,
Higgs or triple gauge couplings), or beyond it (once new interactions have been
discovered). A precise cross-section prediction is therefore needed.
For those processes and for the Drell-Yan process $pp \ra \gamma^* X$,
the NNLO corrections have been recently been calculated \cite{Anastasiou}
for Tevatron and LHC energies. The NNLO cross-section is a bit smaller 
in NLO, but it remains within the NLO uncertainties.
The residual scale dependence is reduced to $1\%$ in NNLO and the shape of the rapidity distribution
of the bosons is unchanged. Therefore, it will probably be possible
to simulate these processes with NLO QCD programs like MC@NLO and apply a normalisation
factor ($K$-factor). One has to see, however, if this conclusion still holds once
the detector acceptance is folded in. Together with the
recent calculations of the NNLO DGLAP splitting functions \cite{nnlo} 
these calculations will allow
to include these data in future NNLO QCD fits to HERA, Tevatron and LHC data.

\subsection{Beauty and charm production}

One area of much recent phenomenological activity where HERA and
Tevatron data have both played an important role is the question of
how charm and beauty are produced in hadronic collisions. It is
obviously important to understand these processes for LHC, since
b-tagging is a vital tool in many searches. After an interesting
history in which some large discrepancies were reported, the latest
data on beauty production at HERA and Tevatron are reasonably well
described by perturbative QCD (see \cite{excess} for an entertaining
and informative review). Nevertheless, this is often within fairly
large uncertainties, and more precise data from HERA II, in particular
using the improved tagging capabilities of the experiments' new vertex
detectors (demonstrated by H1 at this meeting~\cite{h1vxd}) should be
able to provide further and more stringent tests of the phenomenology.

\section{Testing ground for non- or semi-perturbative models}
 
The almost on-shell photons which come along with the electron beam at
HERA collide with protons, and these photons can fluctuate to acquire
a hadron-like structure. Therefore HERA can look like a hadron-hadron
machine ({\it i.e} hadronic photon vs proton), but can also do ``simpler"
measurements with a pointlike photon (for example in DIS, or direct
photoproduction). This ability to turn on and off the hadronic nature
of the photon gives HERA unique handle for testing non- and
semi-perturbative models of remnant-remnant interactions.

Such interactions are an inevitable property of hadronic collisions,
and have an impact on jet energies and profiles, energy flow and the
isolation of photons. They are a natural consequence of eikonalisation
of the parton model in high density PDF region~\cite{jimmy}, and as
such are also responsible for unitarising the total cross-section.
They are also related to diffractive factorisation breaking and
rapidity gap survival probability, as well as to
absorption/rescattering corrections to forward proton and neutron
production.

\subsection{The underlying event and minimum bias physics}

Multiparton interaction models~\cite{jimmy,zijl,phojet} have been
shown~\cite{jetweb,rick,craig} to give the best description of the
final state in $p\bar{p}$ and $\gamma p$ interactions, and also do
well in $\gamma\gamma \rightarrow {\rm jets}$.  The extrapolation of
such models to the LHC inevitably involves very large
uncertainties. 
Studying data from several experiments and
processes at existing experiments allows us to learn about the energy
dependence and target particle dependence of models and at least
reduce the ``phase-space'' of possible models. 

\begin{figure}[!thb]
\vspace*{4.6cm}
\begin{center}
\includegraphics{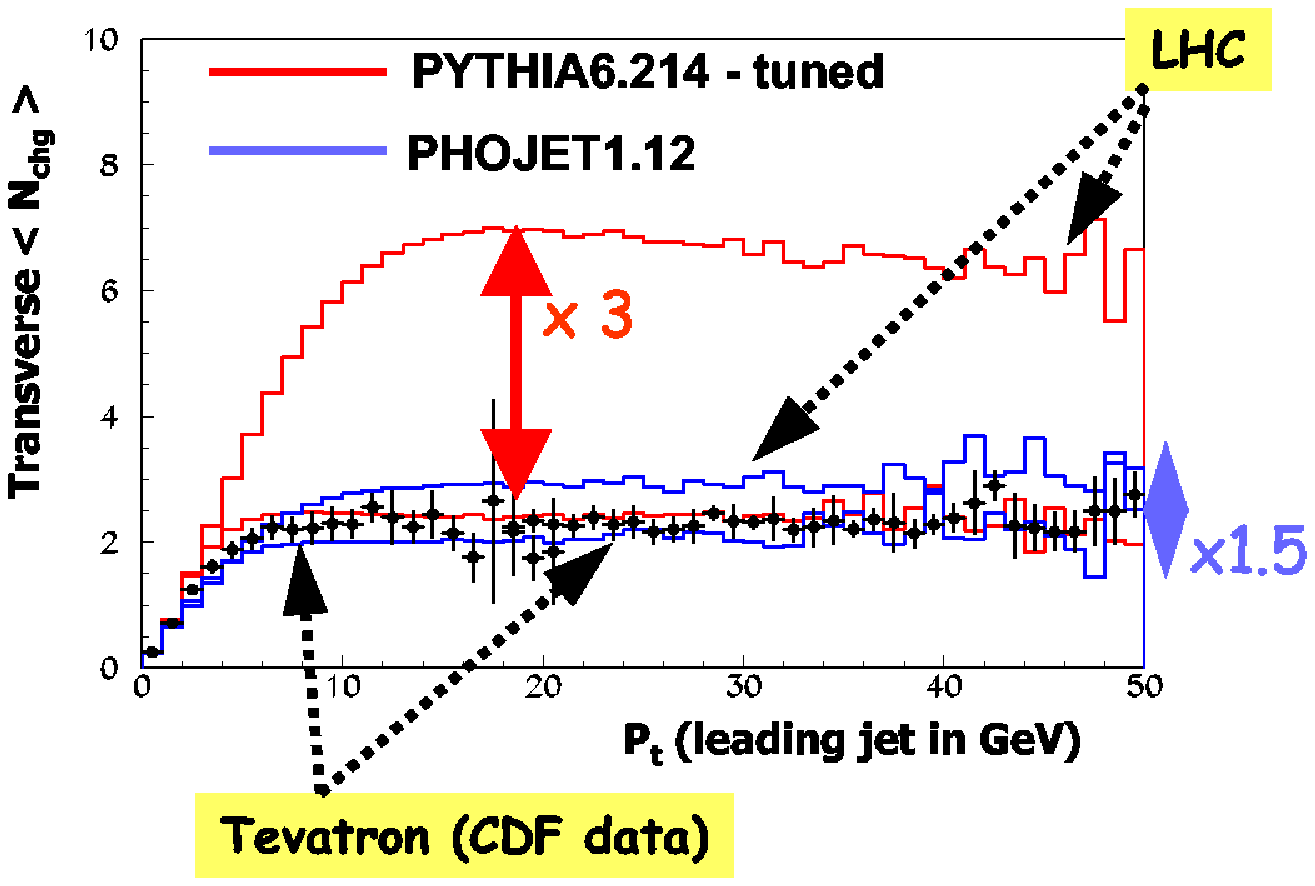}
\includegraphics{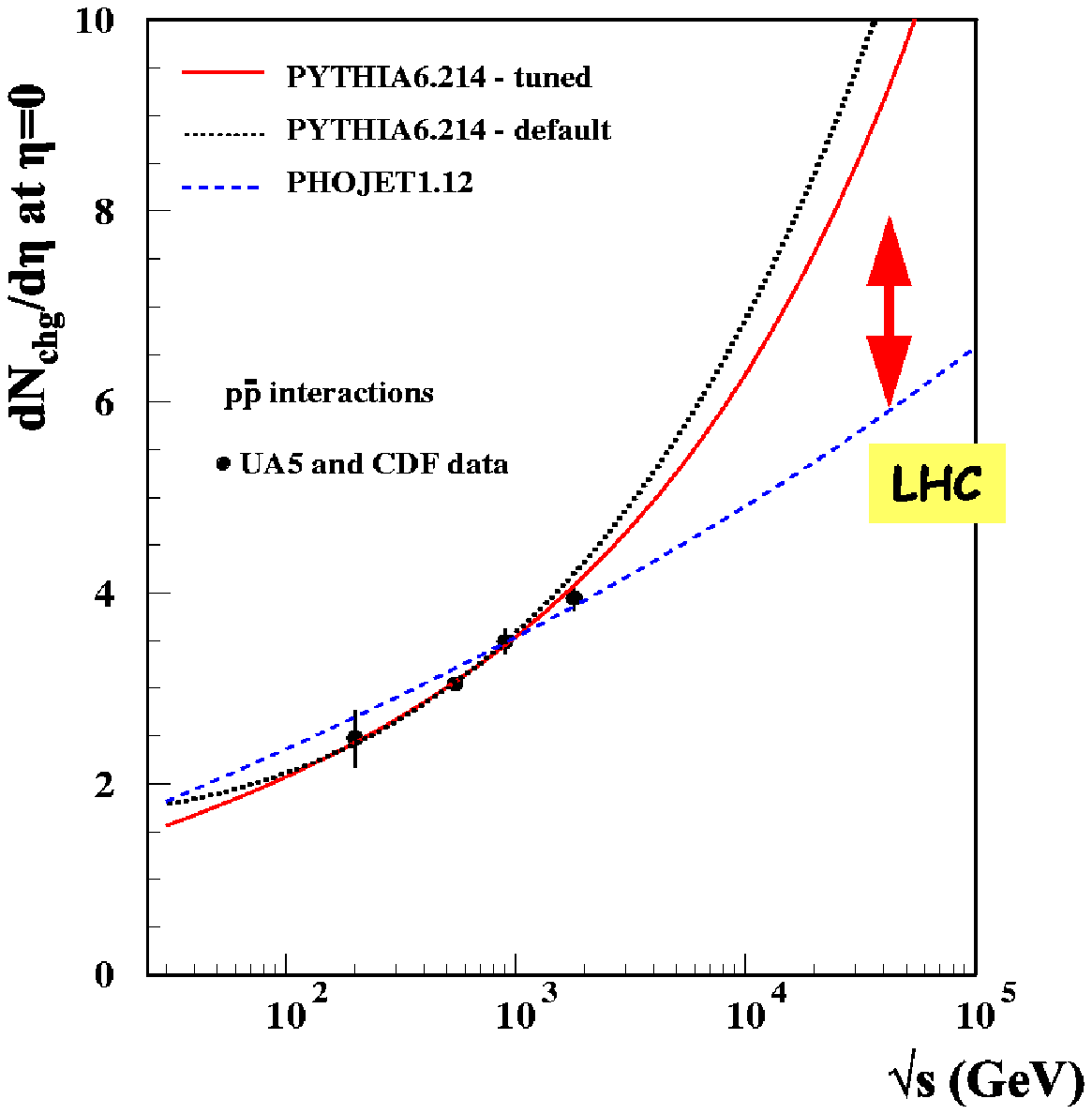}
\begin{picture}(0,0)
\put(-165,1){a)} 
\put(  30,1){b)} 
\end{picture}
\caption[*]{a) Average charged particle multiplicity in the hemisphere
not containing the hard jets as a function of the leading jet
transverse momentum.  b) Average charged particle multiplicity in the
central rapidity region as a function of the $\sqrt{s}$. Figure taken
from \cite{moraes}.
\label{fig:minbias}
}
\end{center}
\end{figure}

As an illustration, the average charged particle multiplicity
in the event hemisphere not containing the hard jets in a
$p p$ collision is shown as a function of the leading jet transverse
energy in Fig.~\ref{fig:minbias}a. Shown is the prediction from 
a multi-parton interaction model implemented in PYTHIA and
from a multi-pomeron exchange model based on the dual parton model
for soft and semi-hard particles as implemented in PHOJET \cite{phojet}.
The PYTHIA model was tuned to a variety of hadronic final state
data at SPS and Tevatron \cite{craig}. For PHOJET no tuning
was needed. While both models describe the data at low $\sqrt{s}$, 
they give predictions which differ by a factor of $3$ at LHC.
The dependence of the charged particle multiplicity at the central
rapidity is illustrated in Fig.~\ref{fig:minbias}b. Large differences
are found in the predictions at high $\sqrt{s}$.

Early data from the LHC
will of course be the best way to improve the models; however, even
this is not fool-proof, as the correlations between the hard
subprocess and the underlying event are non-trivial, meaning that
minimum bias events, while useful, are not a 100\% reliable guide.  It
remains important to have a physically consistent and well-motivated
model which is tuned to the widest selection of relevant data.  The
impact of uncertainties on, for example, the minijet vetoes proposed
for identifying vector-boson fusion processes~\cite{mjv,atlastdr} is
large~\cite{ww}.

New ideas to describe soft hadron collisions and the interplay between
soft and hard contributions, developed at HERA, might help to develop
new and better models.  Such models could make use of
$k_T$-factorisation, of skewed parton distribution describing the
correlation of partons at different momentum fraction in the protons
or of the dipole model to describe the transition between soft and
hard scattering.

Another area where HERA data are relevant is in the determination of
the proton PDFs at low x, since these are an important input to
multiparton interaction models. In fact, present tunings of the
minimum bias event models are only valid for a given PDF.
If the $\ptmin$ for hard scatters is
around $1.0-3.0$~{\rm GeV} or so, the range of values favoured by several
of the studies mentioned above, then the models are sensitive to
values of x down to $5 \times 10^{-6}$. This is a region where the
effects of saturation (gluon recombination) may be important.

Possibilities which have not been fully explored yet with HERA data
include detailed studies of the behaviour of jet-finding algorithms
in the same $\ETJ$ and $\ETAJ$ region both with without an underlying
event. It may also be possible to test models which predict both
minimum bias and underlying event by using the electron-tagged
photoproduction samples to obtain ``zero-bias'' events.

\subsection{Leading baryon production}

The physics which leads to the enhanced activity in the central region
known as the underlying event is closely related to the rescattering
(or absorptive) effects which have an impact of leading baryon rates.
In models for these processes, a leading proton or neutron is
initially present due to a low-$t$ pomeron or pion exchange, but is
rescattered and destroyed by remnant-remnant interactions. Such
effects can be rather directly investigated at HERA by appealing once
again to the ability to switch between pointlike (in DIS or direct
photoproduction) photons and the hadronic photons which dominate the
photoproduction cross-section.

The leading neutron energy spectrum at HERA has been shown to be well
described by pion exchange. In inclusive
photoproduction~\cite{fnczeus} there is no hard scale, the photon is
dominantly hadronic,and the forward neutron rate is expected to be
affected by rescattering. However, in DIS~\cite{fnczeus,fnch1}, there
is a hard scale present. In addition, the photon is pointlike and
therefore has no remnant to undergo rescattering. The forward neutron
rate is correspondingly higher. In charm
photoproduction~\cite{dstarfnc}, there is again a hard scale, provided
by the charm mass. Some contribution from the hadronic photon is
expected to be present, but this is suppressed with respect to the
inclusive case, at least for inclusive dijet charm
events~\cite{dijc}. There is no evidence for rescattering in these
events, with the measured neutron fraction of $9 \pm 1\%$ being in
good agreement with the DIS rate, and inconsistent with the rate for
inclusive photoproduction. Finally one can consider dijet
photoproduction. Here a hard scale is present, but one can select
between hadronic and pointlike photons using the $\xgo$
variable~\cite{dij}. There is no acceptance-corrected measurement of
the ratio available as a function of $\xgo$, but both ZEUS and H1
uncorrected data exhibit a trend which indicates that absorptive
effects are reduced as $\xgo$ increases~\cite{dijfnc}.

\subsection{Diffractive particle production as a search channel}

One major source of interest in the measurements of forward neutron
production, discussed above, is their relation to rates for leading
proton production.  This is an area of increasing interest for LHC
experimentalists, and there as been much phenomenological progress in
the past year or so, reflected by several talks in the diffractive
session at the workshop~\cite{lplhc}. This may be the cleanest way to
discover a low-mass Higgs at the LHC, and other exotic neutral
particles may also be accessible. Experimentally the technique
requires leading-proton tagging to provide excellent resolution on the
energy of the central system. A proton spectrometer designed to
achieve this must be triggered in conjunction with the central
detector, and would also do some excellent diffractive QCD
physics. The phenomenological predictions require a good understanding
of diffractive processes, particularly diffractive PDFs and
factorization breaking. 

At HERA, along the lines of the leading neutron discussion in the
previous section, one expects the leading proton (and associated
rapidity gap) to be destroyed by rescattering, with a probability
which is higher for hadronic photons than for pointlike photons. Dijet
photoproduction in association with a forward rapidity gap has been
measured~\cite{diffdij} and compared to LO Monte Carlo models as well
as NLO QCD calculations~\cite{kk}. Interestingly, the LO Monte
Carlo's, which in this case do not include any remnant-remnant
interactions, describe the data well without any need for a
rescattering correction.  However, in the NLO calculations, agreement
is only seen if a rescattering correction of around
0.34~\cite{kaidalov} is applied to the hadronic photon component.

If the substantial investment of effort and money required for leading
proton spectrometry is to be made, it is obviously vital to gain as
much confidence as possible in the cross-section predictions for LHC.
Demonstrating a phenomenology which can accurately accommodate the
rescattering effects in leading proton and leading neutron production
in photoproduction and DIS, as well as describing underlying events,
would be a major step towards this. Away from HERA, crucial input is
also required from the from measurements at the Tevatron of
diffractive jet and particle production~\cite{tev}.

\section{Summary}

HERA is a great lab for testing the standard model, particularly QCD.
Good data are already available on the hadroproduction of jets, of
photons and of rapidity gaps. There is further precise heavy flavour
data to come from the HERA II program. Systematic efforts to make best
use of this data are underway and should intensify.

Uncertainties from QCD effects are expected to be the limiting factor in
many key measurements and searches at the LHC. Working out what we
need to know from current colliders should be a priority now for those
interested in LHC data, while new measurements can still be proposed.

\section*{Acknowledgements} We thank the DIS04 organisers
for organising a good meeting in a nice place.  The reader is referred
to the current CERN/DESY workshop on HERA and the LHC~\cite{heralhc},
and the authors gratefully acknowledge useful discussions with their
fellow workshop participants both there and in Ko\v sice.

\end{document}

%% file: jmbsym.tex
\newcommand{\newc}{\newcommand}

\newc{\Lumi}{{\cal L}}

\newc{\ra}{\rightarrow}
\newc{\Ra}{\Rightarrow}

\newc{\pom}  {I\hspace{-0.2em}P}

\newc{\gev}{{\rm GeV}}

\newc{\rpv}{{\not \!\! R_p}}
\newc{\rpvm}{{\not  R_p}}

\newc{\gsim}{{\stackrel{>}{\sim}}}
\newc{\lsim}{{\stackrel{<}{\sim}}}
\newc{\sleq} {\raisebox{-.6ex}{${\textstyle\stackrel{<}{\sim}}$}}
\newc{\sgeq} {\raisebox{-.6ex}{${\textstyle\stackrel{>}{\sim}}$}}

\def\3{\ss}

\newc{\ETJ}{E^{{\rm jet}}_T}

\def\xgo{x_\gamma^{\rm OBS}}

\def\ETAJ{\eta^{\rm jet}}

\def\ptmin{\hat{p}_T^{\rm min}}

\def\Mt{M_{\rm top}}

\def\q2{{\rm Q}^2}
\def\p2{{\rm P}^2}
\def\gev2{{\rm GeV}^{2}}
\def\F2g{F_2^\gamma}
\def\f2c{F_2^{\rm charm}}

\def\d0{D^{0}}

\def\begr{\begin{flushright}}
\def\endr{\end{flushright}}

\def\begl{\begin{flushleft}}
\def\endl{\end{flushleft}}